\documentclass[aps,prd,twocolumn,amsmath,superscriptaddress,preprintnumbers,amssymb,showpacs,floatfix,nofootinbib,longbibliography]{revtex4-1}

\usepackage{graphicx,dcolumn,hyperref,xcolor,bm,natbib}
\usepackage{calligra}
\usepackage{egothic}
\usepackage[T1]{fontenc}
\newfont{\rsfsten}{rsfs10 scaled 1200}
\newfont{\rsfsseven}{rsfs10 scaled 1200}
\newfont{\rsfsfive}{rsfs10 scaled 1200}
\usepackage{epsfig}
\usepackage{units, fontawesome}

\newcommand{\aap}{{Astron.~Astrophys.}}

\newcommand{\mnras}{{Mon.~Not.~R.~Astron.~Soc.}}

\def\lsim{\mathrel{\raise.3ex\hbox{$<$\kern-.75em\lower1ex\hbox{$\sim$}}}}
\def\gsim{\mathrel{\raise.3ex\hbox{$>$\kern-.75em\lower1ex\hbox{$\sim$}}}}

      \newcommand{\cL}{{\cal L}}   \newcommand{\cO}{{\cal O}}

\newcommand{\beq}{\begin{equation}} \newcommand{\eeq}{\end{equation}}

\begin{document}

\title{
A Phantom Menace: On the Morphology of the Galactic Center Excess
}

\author{Samuel D.~McDermott}
\affiliation{Fermi National Accelerator Laboratory, Batavia, Illinois, 60510, USA}
%\email{sammcd00@fnal.gov, ORCID: orcid.org/0000-0001-5513-1938}
\author{Yi-Ming Zhong}
\affiliation{Kavli Institute for Cosmological Physics, University of Chicago, Chicago, IL 60637, USA}
%\email{ymzhong@kicp.uchicago.edu, ORCID: orcid.org/0000-000X-XXXX-XXXX}
\author{Ilias Cholis}
%\email{cholis@oakland.edu, ORCID: orcid.org/0000-0002-3805-6478}
\affiliation{Department of Physics, Oakland University, Rochester, Michigan, 48309, USA}

\date{\today}

\begin{abstract}
The characteristics of the Galactic Center Excess (GCE) emission observed in gamma-ray energies -- especially the morphology of the GCE -- remain a hotly debated subject. The manner in which the dominant diffuse gamma-ray background is modeled has been claimed to have a determining effect on the preferred morphology. In this work, we compare two distinct approaches to the galactic diffuse gamma-ray emission background: the first approach models this emission through templates calculated from a sequence of well-defined astrophysical assumptions, while the second approach divides surrogates for the background gamma-ray emission into cylindrical galactocentric rings with free independent normalizations. At the latitudes that we focus on, we find that the former approach works better, and that the overall best fit is obtained for an astrophysically motivated fit when the GCE follows the morphology expected of dark matter annihilation. Quantitatively, the improvement compared to the best ring-based fits is roughly 6500 in the $\chi^2$ and roughly 4000 in the log of the Bayesian evidence.
\href{https://github.com/samueldmcdermott/gce}{\faGithub}  
\end{abstract}

\preprint{FERMILAB-PUB-22-650-T}

\maketitle

\noindent {\bf Introduction:}
For over a decade, claims have persisted that a surfeit of photons are present in \textit{Fermi} Large Area Telescope (\textit{Fermi}-LAT) \cite{Gehrels:1999ri, fermiURL} observations. This so-called Galactic Center Excess (GCE) emission has been observed around the center of the Milky Way at GeV energies \cite{Goodenough:2009gk, Hooper:2010mq, Abazajian:2010zy, Hooper:2011ti, Hooper:2013rwa, Gordon:2013vta, Abazajian:2014fta, Daylan:2014rsa, Calore:2014xka, Zhou:2014lva, TheFermi-LAT:2015kwa, Linden:2016rcf, Cholis:2021rpp}. Details of this emission appear to depend on high-level choices, such as the region of interest (ROI) considered and how the astrophysical emission is modeled, but the existence and the spectrum of the GCE appear very stable, despite the fact that our knowledge of the point-source catalog has dramatically increased in completeness since the discovery of the GCE \cite{Zhong:2019ycb, Cholis:2021rpp}. The central question we wish to address in this work is whether or not the morphology of the GCE is roughly spherically symmetric \cite{Daylan:2014rsa, Calore:2014xka, TheFermi-LAT:2015kwa, DiMauro:2021raz, Cholis:2021rpp} or follows the morphology of stellar populations in the Galaxy \cite{Macias:2016nev, Bartels:2017vsx, Macias:2019omb, Abazajian:2020tww, Calore:2021jvg, Pohl:2022hydrogen}. This is motivated by the question of whether the GCE is produced by dark matter (DM) annihilation, which is expected to be approximately spherically symmetric and could produce GeV emission, or by millisecond pulsars, whose spatial distribution may correlate with stellar populations and whose energy spectrum peaks in intensity around a GeV.

Given the large number of independent degrees of freedom across the sky (equal to the number of unmasked pixels across all energy bins), arriving at an expected background model necessitates many assumptions. Works such as Ref.~\cite{Cholis:2021rpp} assume that the inner galaxy is dominated by galactic diffuse emission originating from nearly steady-state astrophysical processes. These templates are modeled by choosing a set of well-defined astrophysical assumptions controlled by a number of ``hyperparameters'' describing how cosmic rays are injected, propagate, and interact with the interstellar medium to produce gamma rays. Uncertainties on those hyperparameters are accounted for by creating a large number of models of the interstellar medium and the cosmic-ray sources of the inner galaxy, and also by allowing for some normalization freedom between different astrophysical mechanisms for the entire ROI after the diffuse modeling is complete. This freedom on the hyperparameters of the diffuse modeling and on the resulting independent normalizations account for the uncertainties in the underlying hypotheses. The \textit{Fermi} bubbles, a prominent emission component that is known to be non-steady state, are included independently. 

Modeling the sky without these astrophysical modeling assumptions necessarily introduces a different number of fit parameters. The approach in \cite{Macias:2016nev, Macias:2019omb, Abazajian:2020tww, Pohl:2022hydrogen}, following official {\it Fermi} collaboration analyses \cite{TheFermi-LAT:2015kwa, Fermi-LAT:2016zaq}, introduces cylindrical, galactocentric templates that, in projection, look like rings, and which we will refer to henceforth as ``ring-based templates''. These rely on gas maps and inverse Compton scattering maps, plus a set of two ``residual'' components, leading to 16 separate normalizations per energy-bin used, as described below. Here, we perform a comprehensive fit to the ring-based templates from \cite{Pohl:2022hydrogen} in order to compare the fit quality to the results of Ref.~\cite{Cholis:2021rpp} in a unified statistical framework.

~\\ \noindent {\bf Statistical Procedure:}
In this work, we compare the results of the astrophysically motivated templates of \cite{Cholis:2021rpp} to the approach of \cite{Macias:2016nev, Macias:2019omb, Abazajian:2020tww, Abazajian:2020tww, Pohl:2022hydrogen} for background emission. This latter approach includes 16 independent galactocentric cylinders based on expected tracers of gamma ray emission: four rings follow the neutral atomic hydrogen ({\sf HI}) density, four follow the neutral molecular ({\sf H2}) density, six follow the calculated inverse Compton scattering ({\sf ICS}) emission, and two ``residual'' components (one negative valued and one positive valued template, tuned by hand) are included to ensure a good fit. The {\sf HI} and {\sf H2}  rings are (annular) cylinders with boundaries at 3.5 kpc, 8 kpc, 10 kpc, and 50 kpc from the Galactic center. The {\sf ICS} rings are the same, except that the first ring is subdivided at radii of 1.5 and 2.5 kpc as well as at 3.5 kpc; these three divisions are depicted in the Appendix. In addition, we include the isotropic background and a \textit{Fermi} bubbles template. Despite the uncertain low-latitude nature of the \textit{Fermi} bubbles, their morphology in the ROI is fairly well constrained. For our main results, we use maps derived from \cite{Pohl:2022hydrogen}, but for completeness, we report results based on the maps of \cite{Macias:2016nev, Abazajian:2020tww} in the Appendix. In all cases, we have projected these maps (supplied by the authors of \cite{Macias:2016nev, Pohl:2022hydrogen}) onto the Cartesian grid near the Galactic center, accounting for unequal solid angles, subdivided into $0.1^\circ \times 0.1^\circ$ pixels, and smoothed by the energy-dependent \textit{Fermi}-LAT point spread function.

Alongside these rings, we will add candidate excess templates with different spatial distributions that have been suggested to provide good fits to the GCE. Candidate excess morphologies are as follows. First, we consider a contracted, squared, and integrated Navarro-Frenk-White profile \cite{Navarro:1995iw,Navarro:1996gj,Wang:2019ftp} with inner slope 1.2, which follows the expected contribution of DM annihilation to the gamma-ray sky. Next, we consider various stellar populations, which may trace star-forming regions that include significant athermal gamma-ray emission or may harbor significant populations of millisecond pulsars. The stellar populations we test in this work are the boxy bulge (BB) which traces red-clump giants \cite{2002A&A...384..112L, Macias:2016nev, Bartels:2017vsx}, the X-shaped bulge \cite{Macias:2016nev, Bartels:2017vsx, Macias:2019omb}, and a combination of the boxy bulge along with the nuclear stellar cluster and the nuclear stellar disk (${\rm BB^+}$) \cite{2002A&A...384..112L}, normalized as in \cite{Cholis:2021rpp}. 

Our log-likelihood in a given energy bin $j$ is
\begin{equation}
\! \! \! - 2 \ln \mathcal{L}_j \! = \!
2  \sum_{p} \left[\mathcal C_{j,p}\! + \ln(\mathcal D_{j,p}!) \! - \! \mathcal D_{j,p} \ln \mathcal C_{j,p}  \right]  +  \chi^2_{\rm ext}
\label{eq:Likelihood}
\end{equation}
where: $p$ are the unmasked pixels; the expected counts $\mathcal C_{j,p} = \mathcal E_{j,p} \sum_i c^i_j \Phi^i_{j,p} $ are obtained from summing over the astrophysical templates $\Phi^i_{j,p}$ of \cite{Cholis:2021rpp} or \cite{Pohl:2022hydrogen}, plus the \textit{Fermi} bubbles and isotropic templates, plus zero, one, or two excess templates, each of which is multiplied by an independent normalization ($c_j^i$) and by the exposure $\mathcal E_{j,p}$; the data $\mathcal D_{j,p}$ are the same data as in \cite{Cholis:2021rpp}, which we describe again in the Appendix for ease of reference; and $\chi^2_{\rm ext} = [(c^{\rm bub}_j - 1)/\sigma^{\rm bub}_j]^2 + [(c^{\rm iso}_j - 1)/\sigma^{\rm iso}_j]^2 $ is an ``external $\chi^2$'' that provides a penalty when the bubbles and isotropic normalizations deviate too much from their spectra measured at high latitudes \cite{Fermi-LAT:2014sfa, Ackermann:2014usa}. The Bayesian evidence in a given energy bin is the integral of the likelihood over the entire prior manifold. For each model, the total log likelihood or the total log evidence are given by summing the values of the log likelihoods or the log of the Bayesian evidences across all energy bins. In total, we will have eighteen, nineteen, or twenty free parameters $c^i_j$ in each energy bin of our fits when fitting with the ring-based templates of \cite{Pohl:2022hydrogen}, corresponding to the cases of no excess emission, a single excess component, or two components simultaneously varying; similarly, we have four or five free parameters $c^i_j$ in each energy bin of our fits when fitting with the templates of \cite{Cholis:2021rpp}.

As in \cite{Cholis:2021rpp}, the ROI is $2^\circ \leq |b|\leq 20^\circ$, $|\ell|<20^\circ$ in galactocentric coordinates. We enforce a mask at $|b|<2^\circ$ for a number of reasons. Most importantly, restricting to higher latitudes allows us to avoid putting statistical weight on the regions of the sky, in particular the Galactic disk, where the astrophysical emission is the brightest. Thus, masking the disk puts statistical weight on regions where the signal to background ratio is highest and where variations between different proposed excess morphologies differ most significantly. Also, masking the Galactic disk omits the contribution of undiscovered point sources in a region where they are expected to be both numerous and difficult to detect. Together, these considerations suggest that masking the disk allows us to better answer the question of the morphology of the GCE. We also mask all of the point sources in the 4FGL-DR2 catalog \cite{Ballet:2020hze} with a mask that is test statistic- and energy-dependent, following \cite{Cholis:2021rpp}. We reproduce the details of this mask in the Appendix for ease of reference.

In order to quantitatively understand all emission components, we reconstruct the {\it full posterior manifold} of each energy bin. The approach described here differs from \cite{Macias:2016nev, Macias:2019omb, Abazajian:2020tww, Abazajian:2020tww, Pohl:2022hydrogen}, which relied on the local optimizer {\tt MINUIT} \cite{James:1994vla} as implemented in {\it Fermi} {\tt ScienceTools}\footnote{\url{https://fermi.gsfc.nasa.gov/ssc/data/analysis/}}. Our approach requires numerical methods that allow us to sample from the posterior manifolds, with the benefit of being able to find a global rather than local optimum.

To ensure good convergence properties for the high-dimensional posteriors of interest, we use two different samplers. First, we use a nested sampler \cite{Skilling:2004nest, Skilling:2006nest, Feroz:2009nest} with the goal of comprehensively exploring the posterior. This is useful because the posterior of interest is high-dimensional, highly correlated, and multimodal. This also allows us to report the Bayesian evidence attributed to each model, which facilitates comparison between models with different numbers of parameters. After finding the region that plausibly contains the maximum likelihood parameters, we use the No-U-Turn Sampler \cite{Hoffman:2011nuts} implementation of Hamiltonian Monte Carlo (HMC) \cite{Neal:1996nn, Neal:2011hmc, Betancourt:2017hmc}, which provides an independent check of the nested sampler and which quickly attains the best possible fit for each model. For our numerical work, we rely on the implementations of the nested sampler in {\tt dynesty}\footnote{\url{https://doi.org/10.5281/zenodo.3348367}} \cite{Speagle:2020dynesty} and the No-U-Turn Sampler in {\tt numpyro}\footnote{\url{https://num.pyro.ai/}} \cite{phan2019composable, bingham2019pyro}, respectively. Our {\tt dynesty} chains terminate with stopping criterion {\tt dlogz=1}. Our {\tt numpyro} runs are initialized from the final point of the {\tt dynesty} chains and are allowed to take $10^4$ steps before stopping. We use very wide, log-flat priors, and observe good convergence. 
In the Appendix, we provide results from tests on the width of the log-flat priors. Our basic conclusions are unaffected by implementing even the most agnostic possible priors on the magnitude of the templates used to model the galactic diffuse emission.

Both of these samplers work well with the number of fit parameters encountered in this work. In future work, especially work that seeks to include the Galactic disk and the large number of point sources therein, HMC may provide unique access to the very high-dimensional posterior. The No-U-Turn Sampler has been shown to maintain its good scaling behavior and provide converged fits with over a thousand parameters \cite{Shen:2021mwhmc}.

~\\ \noindent {\bf Results:}
\begin{table}[t]
\caption{Comparison of models of the GCE. The first seven results, generated in this work, rely on the ring-based method of \cite{Pohl:2022hydrogen} to describe astrophysical emission. The final five results utilize templates from \cite{Cholis:2021rpp}.}
\begin{center}
\begin{tabular}{c|c|l|l}
Excess Model & Bgd.~Templates & $-2 \Delta \! \ln \cL$ & $\Delta \!\ln \mathcal B$ \\ \hline
No Excess & ring-based \cite{Pohl:2022hydrogen} & 0 & 0 \\
X-Shaped Bulge & ring-based \cite{Pohl:2022hydrogen} & $+30$ & $-190$ \\
Dark Matter & ring-based \cite{Pohl:2022hydrogen} & $-237$ & $+12$ \\
Boxy \& X-Shaped Bulges & ring-based \cite{Pohl:2022hydrogen} & $-634$ & $+178$ \\
Boxy Bulge & ring-based \cite{Pohl:2022hydrogen} & $-724$ & $+228$ \\
Boxy Bulge ``plus'' & ring-based \cite{Pohl:2022hydrogen} & $-765$ & $+311$ \\ 
Boxy Bulge ``plus'' \& DM & ring-based \cite{Pohl:2022hydrogen}  & $-817$ & $+316$ \\ \hline
No Excess & astrophysical \cite{Cholis:2021rpp} & $-4539$ & $+2933$ \\
Boxy Bulge & astrophysical \cite{Cholis:2021rpp} & $-6398$ & $+3814$ \\ 
Boxy Bulge ``plus'' & astrophysical \cite{Cholis:2021rpp} & $-6477$ & $+3853$ \\ 
Dark Matter & astrophysical \cite{Cholis:2021rpp} & $-7288$ & $+4268$ \\
Boxy Bulge ``plus'' \& DM & astrophysical \cite{Cholis:2021rpp} & $-7401$ & $+4298$
\end{tabular}
\end{center}
\label{tab:results}
\end{table}%
The results of the fits to the ring-based templates of \cite{Pohl:2022hydrogen} with the six models of the GCE enumerated above are given in Tab.~\ref{tab:results}. We compare against the baseline scenario with no excess at all and also against fits with the templates of \cite{Cholis:2021rpp}. For the templates of \cite{Cholis:2021rpp}, we show results for: no excess, in the first row of the ``astrophysical'' results section of that table; when the GCE follows the Boxy Bulge of \cite{Bartels:2017vsx}, in the second row of the ``astrophysical'' results section; and when the GCE follows the Boxy Bulge with an added nuclear bulge component (Boxy Bulge ``plus'' ) as suggested in \cite{Bartels:2017vsx}, in the third row of the ``astrophysical'' results section. We also show results with the GCE following DM annihilation and when the GCE is a combination of the Boxy Bulge ``plus''  and the DM annihilation morphologies (last two rows of ``astrophysical'' results). These are all fitted in the masked $40^\circ \times 40^\circ$ ROI with the likelihood of Eq.~\ref{eq:Likelihood}. For the combination of the Boxy Bulge ``plus''  and the DM annihilation morphologies, we allow their two normalizations to vary independently.

In the first two columns of Tab.~\ref{tab:results}, we provide the description of the excess model and the type of background model it is fit alongside. In the next column, we provide negative two times the log-likelihood of each model minus the log-likelihood of the model with no excess. A lower negative log-likelihood indicates a better fit; the factor of two is included so that the distribution of these values follows a $\chi^2$ distribution. In the final column of the table, we provide the log of the Bayesian evidence of each model, $\mathcal B$, also called the marginal likelihood. This value is the integral of the likelihood for the model over the entire prior manifold, which provides a complementary (and inherently more Bayesian) way to compare models. A higher Bayesian evidence indicates that a model is more suitable for explaining the data. Each of these values should be understood to have an error bar of size $\sim \cO(10)$, since the stopping criterion for our nested sampling runs was {\tt dlogz = 1} for all 14 energy bins.

We find that the best overall fit to the data using a single excess component is provided by the DM annihilation template when fit alongside the background templates developed in \cite{Cholis:2021rpp}, which are based on astrophysical assumptions. 
As shown in the results with two excess components in \citet{Cholis:2021rpp}, the boxy bulge ``plus'' is subdominant to the dark matter annihilation component, and is only important at energies below 0.7 GeV. A boxy bulge ``plus'' morphology on its own is strongly disfavored compared to a DM morphology when using background templates developed in \citet{Cholis:2021rpp}: the $\chi^2$ increases by $\sim 800$. The critical result of this work is that using the ring-based templates to model the gamma-ray sky dramatically worsens the fit, {\it regardless of how we model the excess emission}: with ring-based templates and a GCE that follows the boxy bulge, the value of the $\chi^2$ increases by roughly 6500, and the log evidence is reduced by roughly 4000. Including the tracer of the nuclear stellar component or the X-shaped bulge alongside the boxy bulge does not appreciably change the fit. Fitting the ring-based templates alongside the DM template degrades the fit even further, compatible with findings in \citet{Macias:2016nev, Bartels:2017vsx, Macias:2019omb}. When we use the ring-based templates of \citet{Pohl:2022hydrogen} we agree with their finding that there is preference for a boxy bulge ``plus'' morphology over purely a DM one. Similar to the case with astrophysically motivated templates, simultaneously including the DM and the boxy bulge ``plus'' templates marginally improves the fit.

Nevertheless, this work in conjunction with \citet{Cholis:2021rpp} makes clear that the relative difference that can be attributed to the excess templates alongside the ring-based emission is dwarfed by the overall fit quality difference due to fitting alongside templates based on astrophysical assumptions. Since \citet{Cholis:2021rpp} found that a GCE with the morphology of DM annihilation provided an improvement in the fit at the level of a $\Delta \chi^2$ of 900 (2750) compared to the boxy bulge (no excess), as reproduced in Tab.~\ref{tab:results}, we conclude that the DM annihilation morphology presently provides the best explanation of the GCE as suggested by \citet{Daylan:2014rsa, Calore:2014xka, TheFermi-LAT:2015kwa, DiMauro:2021raz, Cholis:2021rpp}, though a subdominant contribution from millisecond pulsars to the GCE at energies below 1 GeV has been shown to be still in agreement with the \textit{Fermi} data \citet{Cholis:2021rpp, Zhong:2019ycb}.
We note that in \citet{Cholis:2021rpp}, the preference for a DM morphology over a bulge morphology or no excess at all, was demonstrated for a large number of astrophysical models and not just the best fit one for which we provide specific numbers of $\Delta \chi^2$ here (see Ref.~\citet{Cholis:2021rpp}, for further details). Also for a more extensive discussion on the performance of different bulge morphologies with the various astrophysical templates models see Ref.~\citet{Cholis:2021rpp}.

Investigating the nature of these differences is of paramount importance.
We raise the possibility that the relative fit quality of the excess morphologies is determined predominantly by the astrophysical modeling and only secondarily by inherent morphological characteristics of the excess. This underscores the primacy of the need for high-quality fits to the dominant emission components, and the difficulty in interpreting results without this context.

These differences in measures of goodness of fit are substantial and imply that the current ring-based templates are not flexible enough to reproduce astrophysically self-consistent diffuse emission templates. We think it is an interesting open challenge to produce models with sufficient parametric freedom to form a superset of the astrophysically based models generated in \cite{Cholis:2021rpp}, or to exceed the fit quality of those models. With this much parameter freedom, gradient-aware high-dimensional sampling tools such as HMC will become critically important. Steps in this direction have already been undertaken with parameter freedom attached to every pixel \cite{Storm:2017arh, Bartels:2017vsx} using the {\tt L-BFGS-B} convex optimizer \cite{L-BFGS-B}, which will be interesting to revisit in light of our findings.

\begin{figure}[t]
\begin{center}
\includegraphics[width=0.45\textwidth]{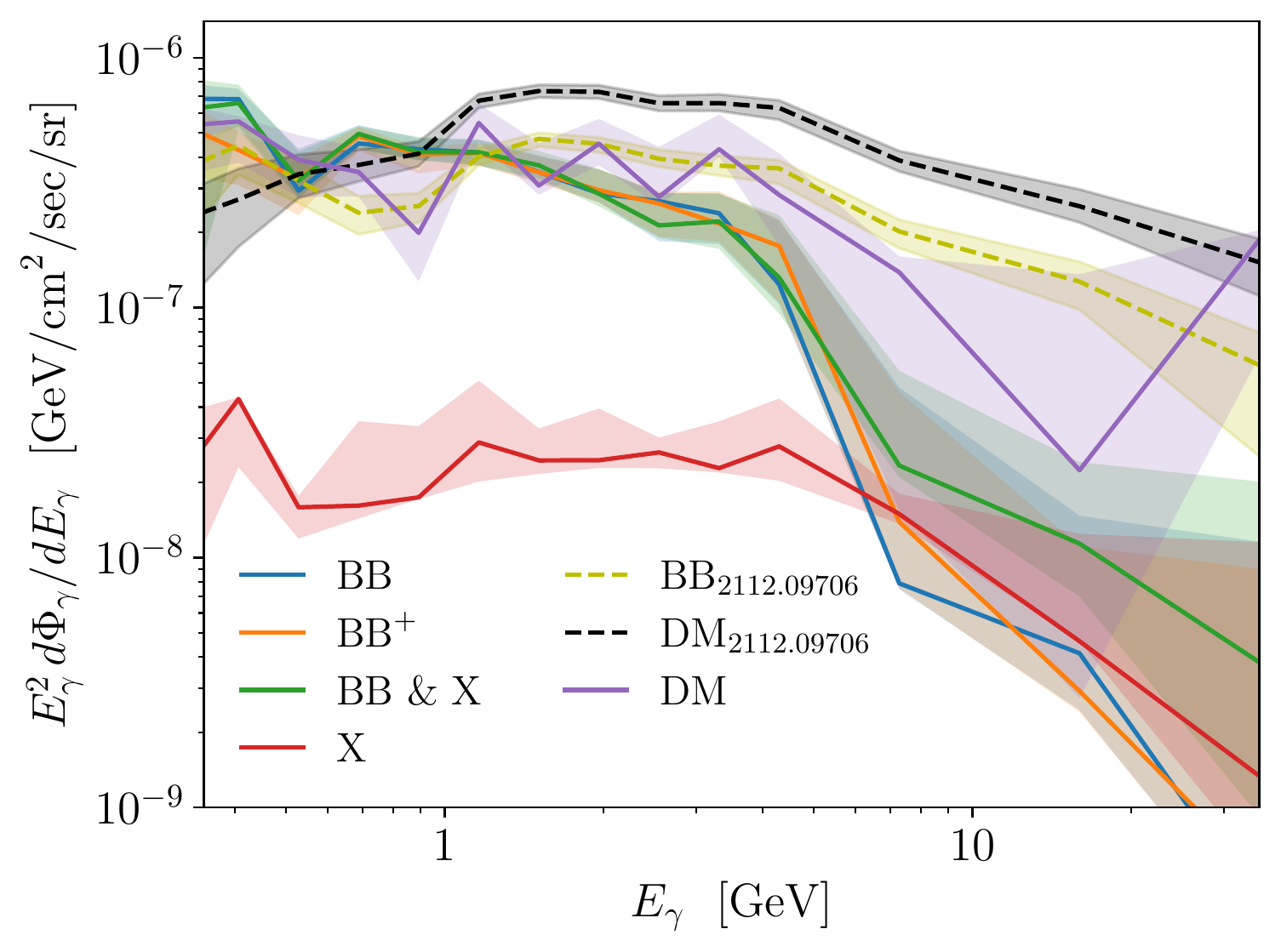}
\caption{Best-fit spectra and 95\% credible intervals of five excesses fit alongside the ring-based templates of \cite{Pohl:2022hydrogen} (solid lines), and two best-fit spectra from \cite{Cholis:2021rpp} (dashed lines).}
\label{fig:spectra}
\end{center}
\end{figure}

In Fig.~\ref{fig:spectra}, we show the best-fit spectra and the 95\% credible intervals for each of the GCE candidates presented in Tab.~\ref{tab:results}, marginalized over the background components of the fit. The solid lines show the fit results obtained in this work alongside the ring-based templates of \cite{Pohl:2022hydrogen}; the dashed lines depict results from \cite{Cholis:2021rpp}. The faintest excess emission is obtained with the X-shaped bulge, which also provides the worst fit to the data. Any excess emission that includes the boxy bulge, including the boxy bulge on its own, the boxy bulge enhanced with the nuclear stellar bulge, and the boxy bulge covarying with the X-shaped bulge, all have very similar spectra. These excess emissions are relatively soft, compatible with having absorbed unmodeled diffuse flux. The DM spectrum with ring-based templates is harder on average than the spectrum arising from these other excesses. It is similar in spectrum to the DM results alongside the astrophysics-based templates from \cite{Cholis:2021rpp}, though lower in flux, and it features ``wiggles'' between energy bins that neither DM annihilation nor millisecond pulsars predict.

The best-fit spectra of many of the background ring-based templates that we fit alongside these excess templates have non-power-law energy-dependent variations that are even more prominent than those visible in the DM spectrum in Fig.~\ref{fig:spectra}. For the {\sf HI} and {\sf H2} rings, the dominant emission is due to the second and third rings in all energy bins, and the non-power-law behavior in energy is only apparent in the subdominant first and fourth rings. Likewise, the negative and positive residuals have relatively smooth, power-law behavior across the entire energy range we study. However, the {\sf ICS} emission is extremely non-smooth in all rings. This behavior is difficult to explain, and it stands in contrast to the smooth astrophysical emission observed in the fits of \cite{Cholis:2021rpp}. For completeness, we provide the best-fit spectra of all of these astrophysical emission components in Fig.~\ref{fig:spectra_r} in the Appendix.

These results conclusively demonstrate that the DM annihilation morphology fit alongside a model based on astrophysical assumptions \cite{Cholis:2021rpp} provides an overwhelmingly better fit to the gamma-ray sky compared to any combination of emission components alongside ring-based models suggested in \cite{Macias:2016nev, Macias:2019omb, Abazajian:2020tww, Pohl:2022hydrogen}. Because we have followed these prior results insofar as possible, the reader may wonder how such a significant difference in conclusions is possible. We stress, however, that our results are compatible with \cite{Macias:2016nev, Macias:2019omb, Pohl:2022hydrogen} in showing a preference for a boxy bulge {\it ceteris paribus} -- the strong preference for the DM annihilation morphology comes only with a fit alongside models based on astrophysical assumptions. We also point out that Ref.~\cite{Cholis:2021rpp} constructs several dozen galactic diffuse emission models, representing different sets of astrophysical assumptions for the conditions in the inner galaxy, each of which provides a better fit for the ROI used in this work compared to the best fit results using the ring-based models of \cite{Macias:2019omb, Pohl:2022hydrogen} for the same ROI. Only a small number of the very worst models considered there prefer a boxy-bulge excess over the DM annihilation excess.

~\\ \noindent {\bf Conclusions:}
In a unified framework, we have shown that the Galactic center excess following a dark matter annihilation morphology fit alongside models based on astrophysical assumptions provides an overwhelmingly better description of the gamma-ray sky outside of the Galactic disk compared to any excess emission component fit alongside ring-based templates, despite their larger number of free parameters.

There are two key differences in our work compared to prior work fitting the gamma-ray data to ring-based regions. First, we use a different region of interest and we take a different approach to point sources: we mask the Galactic disk and the members of the 4FGL-DR2 point source catalog, whereas previous analyses have attempted to model the Galactic disk and the point sources simultaneously with the more distant regions of the Galaxy, which places statistical weight on bright regions of the sky that are poorly understood in gamma-ray energies. Second, we have used modern sampling techniques to ensure convergence to the global best-fit point of the entire parameter space rather than relying on local optimizers that are not adept at finding global optima. Within this unified framework and within the region of interest discussed here, our results demonstrate a strong improvement of fit with a model based on astrophysical assumptions fit alongside an excess with the morphology of dark matter annihilation.

In the future, we think it is important to extend the methods introduced here to the regions of the Galactic disk, where uncertainties on the extended and point source emission are severe and the number of fit parameters will dramatically increase: the 4FGL-DR2 catalog contains of order 1000 point sources and extended sources in the ROI, each of which the Fermi collaboration describes with several constrained (but independent) free parameters. Adding this many free parameters to the fits introduces the ``curse of dimensionality'' to ordinary random-walk MCMC sampling, but gradient-aware techniques such as Hamiltonian Monte Carlo are expected to remain powerful within parameter spaces of these sizes. To facilitate future work with many fit parameters, we are making the code underlying our results using such techniques publicly available\footnote{\url{https://github.com/samueldmcdermott/gce}. The ring-based models underlying these results are based on the original work of \cite{Macias:2016nev, Macias:2019omb, Abazajian:2020tww, Pohl:2022hydrogen}. We are able to share our high-resolution, Cartesian versions of these models upon request.}.

The manner in which we model the gamma-ray background plays a determining role on the conclusions we draw about the subdominant components of the gamma-ray sky, yet we can nevertheless confidently report that dark matter annihilation alongside models based on astrophysical assumptions provides our best current understanding of the gamma-ray sky outside of the Galactic disk. 
Allowing the Galactic center excess to be a combination of two components, a dark matter annihilation profile and a bulge profile, leads to only a minor improvement in the fit compared to the purely dark matter annihilation case, with the bulge-tracing component absorbing only a secondary part of the Galactic center excess emission at energies of <1 GeV, as shown in \cite{Cholis:2021rpp}.
We expect that the techniques introduced in this paper have the promise of taking us closer to resolving the mystery of the Galactic center excess.

\bigskip
\noindent {\bf Acknowledgments:} We thank Andrew Hearin, Dan Hooper, Manoj Kaplinghat, Oscar Macias, Martin Pohl, Nicholas Rodd, Tracy Slatyer and Deheng Song for discussions, and especially Oscar Macias for sharing his templates. We thank Dan Hooper and Oscar Macias for comments on the draft.
Part of this work is performed at the University of Chicago's Research Computing Center. We thank Edward W. Kolb for providing access to the resources. 
SDM thanks Dan Hooper, Cheryl Potts, Booker, and Willett for hospitality while this work was nearing completion.
Fermilab is operated by Fermi Research Alliance, LLC under Contract No.~DE-AC02-07CH11359 with the United States Department of Energy, Office of High Energy Physics. 
YZ acknowledges the Aspen Center for Physics for hospitality during the completion of this work, which is supported by NSF grant PHY-1607611. YZ is supported by the Kavli Institute for Cosmological Physics at the University of Chicago through an endowment from the Kavli Foundation and its founder Fred Kavli.
IC acknowledges that this material is based upon work supported by the U.S. Department of Energy, Office of Science, 
Office of High Energy Physics, under Award No.~DE-SC0022352.

This work would not have been possible without the following open-source code packages:  {\tt numpy} \cite{2011CSE....13b..22V}, {\tt scipy} \cite{2020SciPy-NMeth}, {\tt jupyter-notebook} \cite{9387490}, {\tt ipython} \cite{PER-GRA:2007}, {\tt conda-forge} \cite{conda_forge_community_2015_4774216}, {\tt matplotlib} \cite{2007CSE.....9...90H}, {\tt jax} \cite{jax2018github}, {\tt dynesty} \cite{Speagle:2020dynesty}, and {\tt numpyro} \cite{phan2019composable, bingham2019pyro}.

\begin{appendix}

\section{APPENDIX}

\noindent {\bf Data, Masks, and Rings:}
As in \cite{Cholis:2021rpp}, we use {\it Fermi} Pass 8 data, version P8R3, between 4 Aug 2008 and 14 April 2021 (weeks 9 to 670 of \textit{Fermi}-LAT observations)\footnote{\url{https://fermi.gsfc.nasa.gov/ssc/data/access/}}. We use \textit{Fermi} {\tt ScienceTools P8v27h5b5c8} for selection cuts, exposure-cube files, and exposure maps\footnote{\url{https://fermi.gsfc.nasa.gov/ssc/data/analysis/}}. We use these exposure maps to convert the flux maps to maps of expected counts. We restrict to FRONT-converted {\tt CLEAN} data satisfying {\tt zmax = $100^\circ$}, {\tt DATA$\_$QUAL==1}, {\tt LAT$\_$CONFIG==1}, and {\tt ABS(ROCK$\_$ANGLE) < 52}. All of the data and templates we use are in Cartesian projection and cover galactocentric coordinates in the latitude range $-20^\circ < b < +20^\circ$ and the longitude range $-20^\circ < \ell < +20^\circ$. We bin the sky in pixels of size $0.1^{\circ} \times 0.1^{\circ}$, leading to 160,000 total pixels, although many of these are masked, including the pixels along the Galactic disk below $2^\circ$ in latitude. We account for the unequal solid angle subtended by the pixels by applying a correction factor in latitude.

\setlength{\tabcolsep}{6pt}
\begin{table}[t]
       \caption{The energy bins and the radii of the ``small'' and ``large'' masks, $\theta_{s}$ and $\theta_{l}$, used to mask point sources with test statistic less (greater) than 49, respectively. The last column shows the fraction of pixels masked relative to the total number of pixels in the inner $40^\circ \times 40^\circ$ Galactic center region, including both the 4FGL-DR2 catalog and the Galactic disk.}
    \begin{tabular}{r|rr|c}
            $E_{\textrm{min}}-E_{\textrm{max}} {\rm\,[GeV]}$ &  $\theta_{s} [^{\circ}]$ &  $\theta_{l} [^{\circ}]$ & Masked fraction\\
            \hline 
             $0.275-0.357$ & 1.125 & 3.75 & 71.8\%\\
             $0.357-0.464$ &  0.975 & 3.25 & 62.9\%\\
             $0.464-0.603$ & 0.788 & 2.63 & 52.2\%\\
             $0.603-0.784$ & 0.600 & 2.00 & 38.5\%\\
             $0.784-1.02$ &  0.450 & 1.50 & 29.2\%\\
             $1.02-1.32$ & 0.375 & 1.25 & 23.4\%\\
             $1.32-1.72$ & 0.300 & 1.00 & 19.0\%\\
             $1.72-2.24$ & 0.225 & 0.750 & 16.3\%\\
             $2.24-2.91$ & 0.188 & 0.625 & 13.0\%\\
             $2.91-3.78$ & 0.162 & 0.540 & 12.9\%\\
             $3.78-4.91$ & 0.125 & 0.417 & 11.6\%\\
             $4.91-10.8$ & 0.100 & 0.333 & 11.5\%\\
             $10.8-23.7$ & 0.060 & 0.200 & 10.3\%\\
             $23.7-51.9$ & 0.053 & 0.175 & 10.3\%\\
        \end{tabular}
    \label{tab:PSF_vsE}
\end{table}

The energy binning for the data and the flux maps is as follows. We take photons with $0.275 {\rm \, GeV} \leq E_\gamma \leq 51.9  {\rm \, GeV}$ and subdivide them into 14 energy bins. The first 11 energy bins cover a constant log-width in energy. To ensure that each bin has a similar statistical impact in our fits, the final three energy bins cover a wider log-width in energy, which compensates for the diminishing gamma-ray flux at higher energy. The exact endpoint of each bin is given in the first column of Tab.~\ref{tab:PSF_vsE}. In the final three columns of this table, we also provide details of the point source mask utilized in \cite{Cholis:2021rpp} and in this work. We classify point sources from the 4FGL-DR2 catalog according to their test statistic (TS). We mask point sources with $\text{TS}\leq 49$ ($\text{TS} >49$) with circular masks with radius $\theta_s$ ($\theta_l$), which depend on energy -- any pixel that is within an angular distance of $\theta_s$ ($\theta_l$) contributes nothing to our log-likelihood in that energy bin. The values of these angular parameters are listed in the second (third) column of the table. To give a sense of the impact of this mask, we report the fraction of ROI pixels that are masked (including point sources as well as the Galactic disk) in the final column. At the lowest energies, a majority of pixels are masked, but at a GeV less than a quarter of pixels are dropped, asymptoting to 10\% of pixels at the highest energies.

\begin{figure*}
    \centering
    \includegraphics[width=0.45\textwidth]{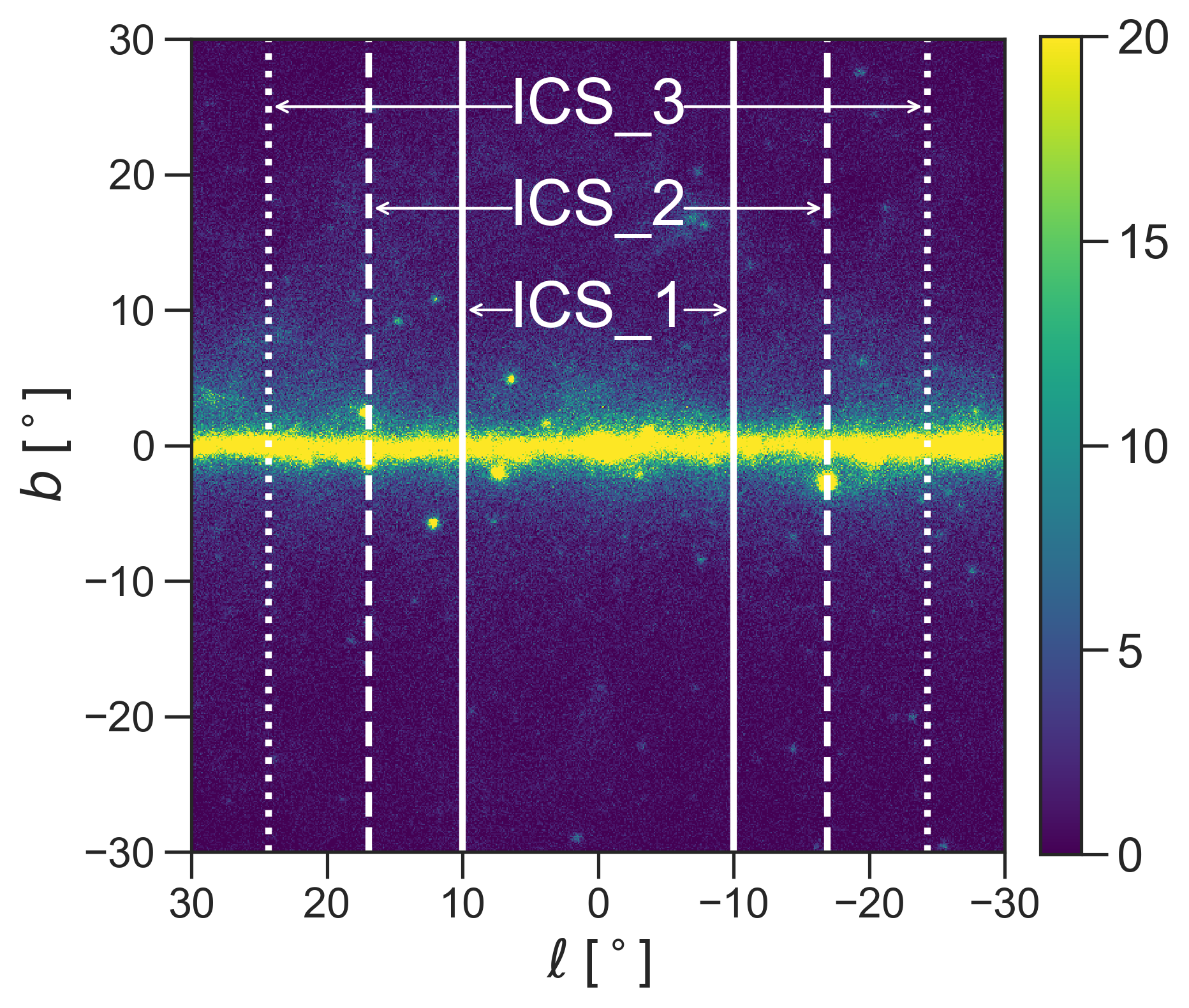}~~
    \includegraphics[width=0.45\textwidth]{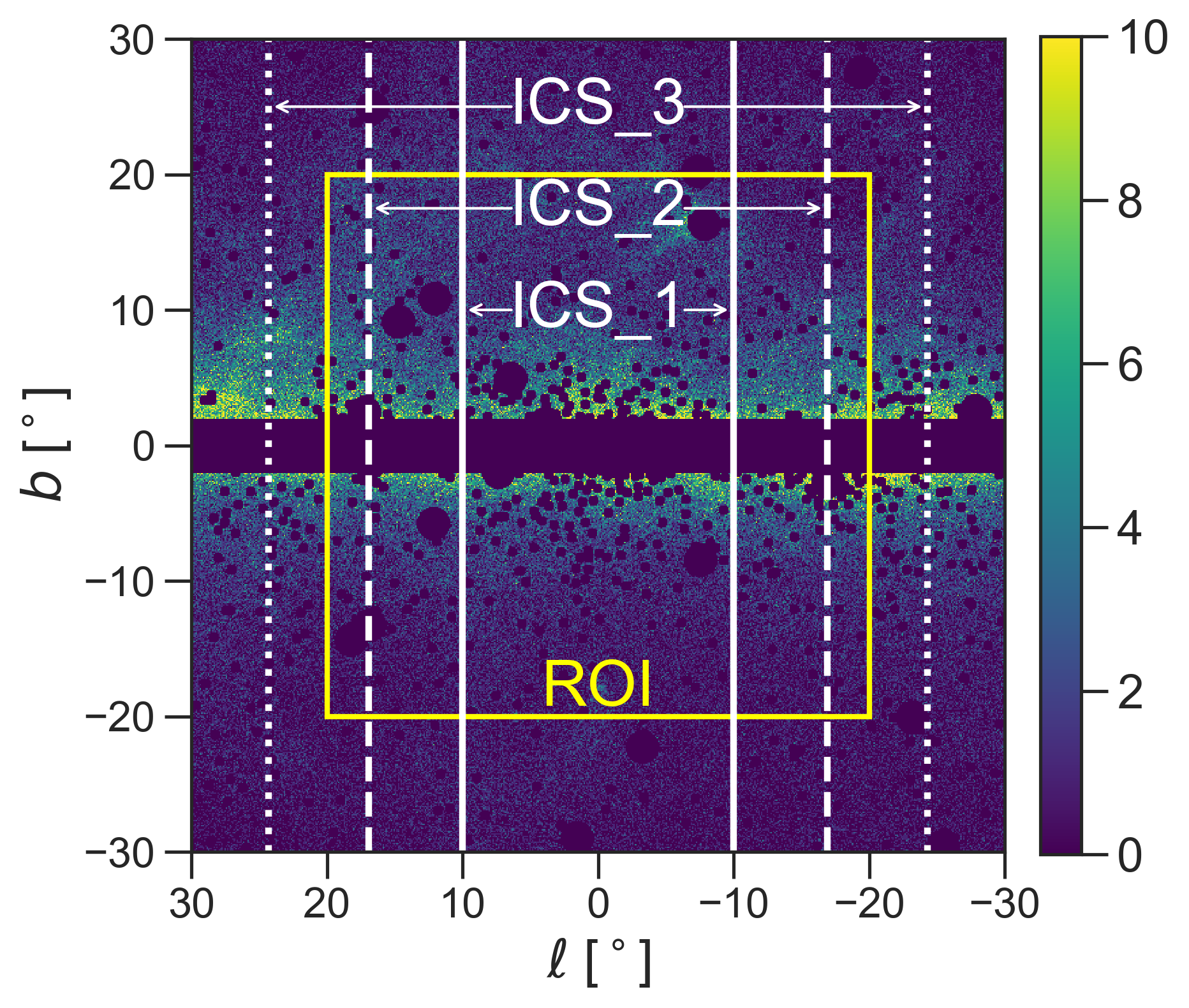}
    \caption{Photons passing our cuts with energy $1.02 {\rm \,GeV} < E_\gamma <  1.32 {\rm \, GeV}$, without (left) and with (right) the mask that we use for our data. For illustration purposes, we show the boundaries of the {\sf ICS\_1}, {\sf ICS\_2}, and {\sf ICS\_3} rings that vary independently in our fits. In the right panel, we show the region of interest in which we perform our fits.}
    \label{fig:data_mask_ring}
\end{figure*}

We show one energy bin of data in Fig.~\ref{fig:data_mask_ring}, expanded to $\pm 30^\circ \times \pm 30^\circ$ for purposes of illustration. This depicts our data in the range $1.02 {\rm \,GeV} < E_\gamma <  1.32 {\rm \, GeV}$. The left (right) panel is the data without (with) the mask described above. We also show the location of the three innermost rings of ICS emission where they intersect the Galactic disk. The outer edge of {\sf ICS\_1} is at 1.5 kpc from the Galactic center, the outer edge of {\sf ICS\_2} is at 2.5 kpc from the Galactic center, and the outer edge of {\sf ICS\_3} is at 3.5 kpc from the Galactic center. Note that, although {\sf ICS\_2} is an annular cylinder with no emission between 1.5 and 2.5 kpc, this ring contributes to emission inside of $-10^\circ \leq \ell \leq 10^\circ$ because of projection effects along the line of sight. The boundary of {\sf ICS\_3} shown at 3.5 kpc is also the outer boundary of the {\sf HI\_1} and {\sf H2\_1} rings that we use. In the right panel, we also show the ROI in which we perform our fits.

~\\ \noindent {\bf Best-Fit Ring-Based Astrophysical Fluxes:}
\begin{figure*}[t]
\begin{center}
\includegraphics[width=0.45\textwidth]{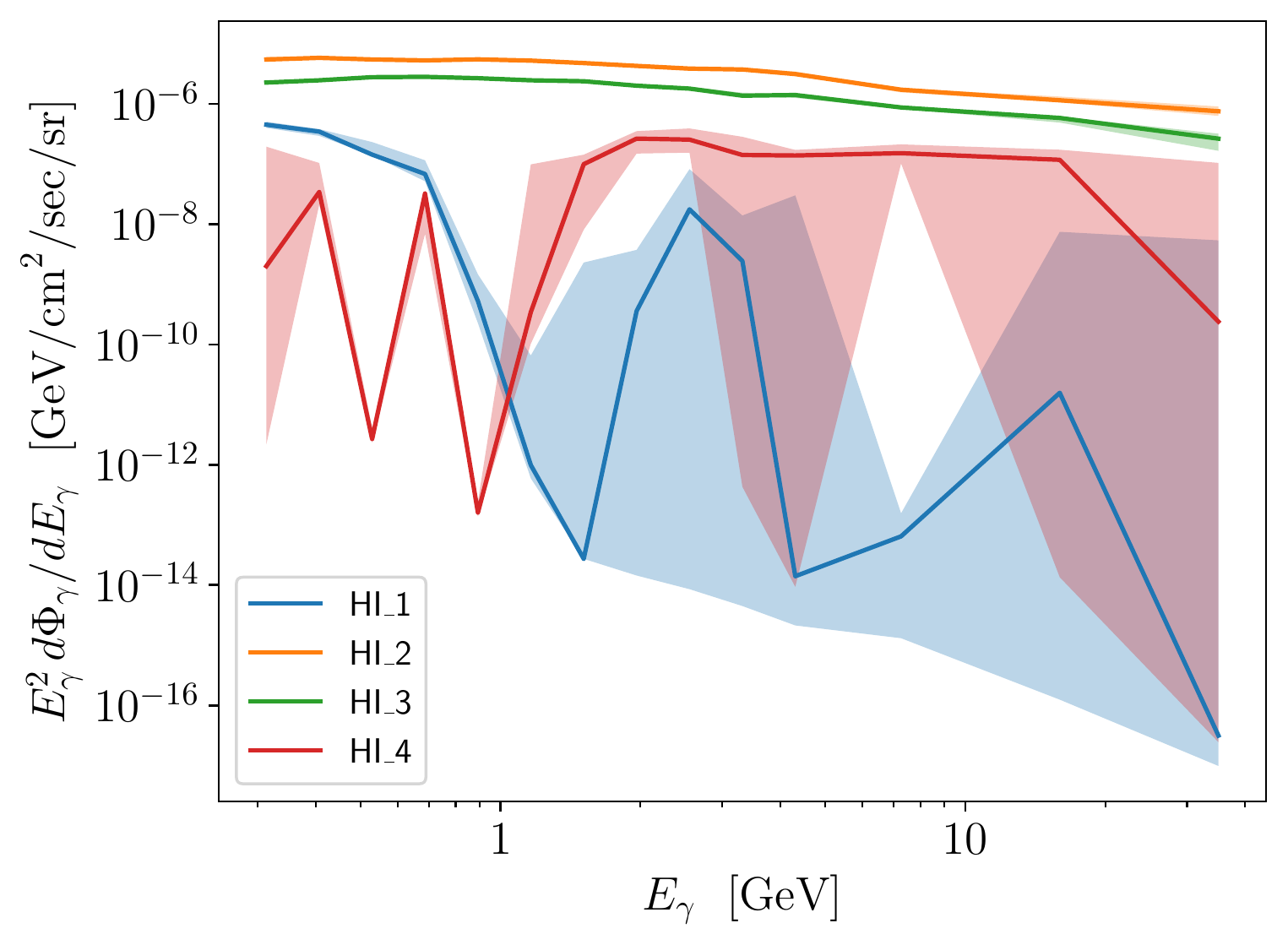}~~
\includegraphics[width=0.45\textwidth]{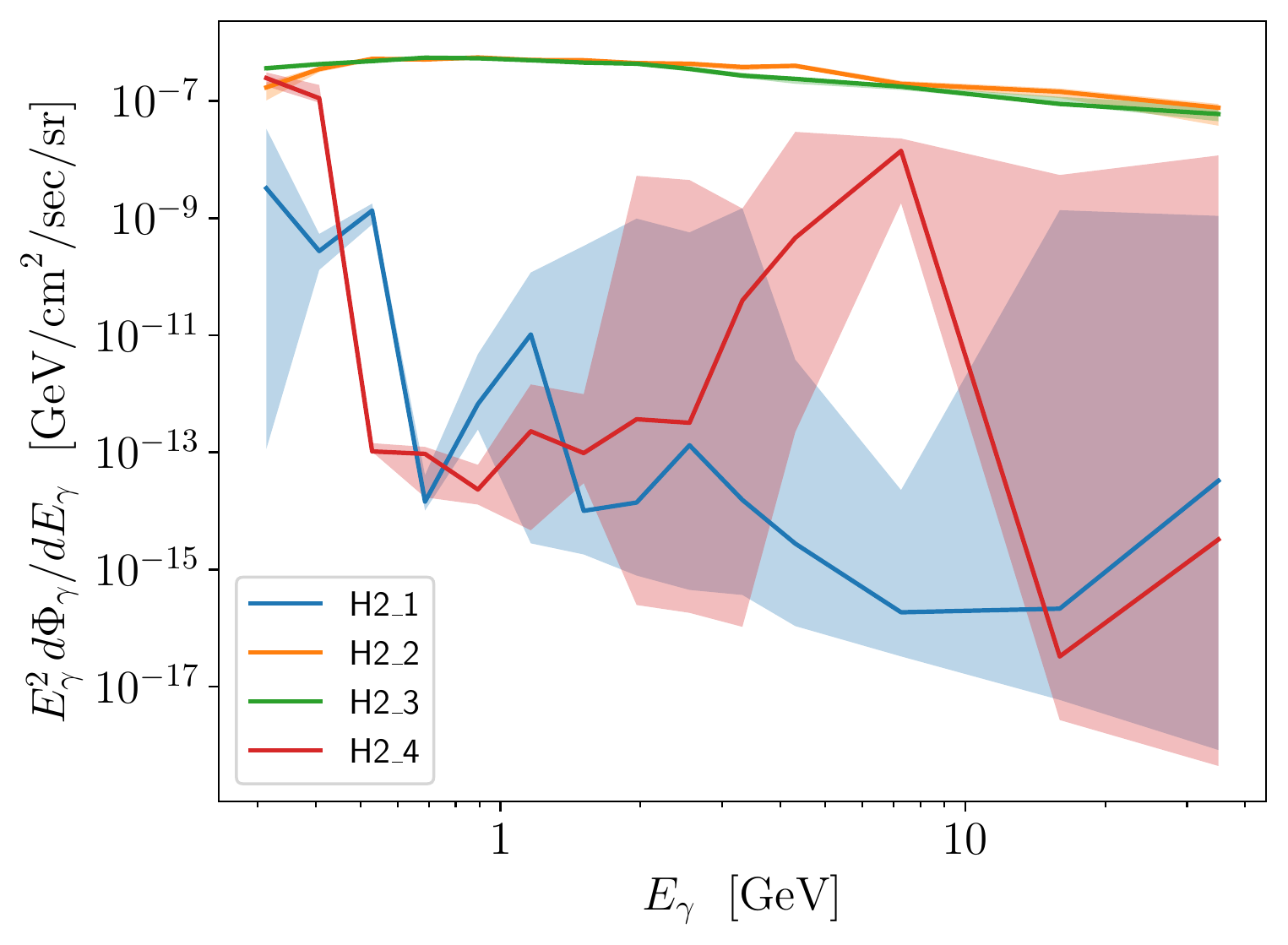}\\
\includegraphics[width=0.45\textwidth]{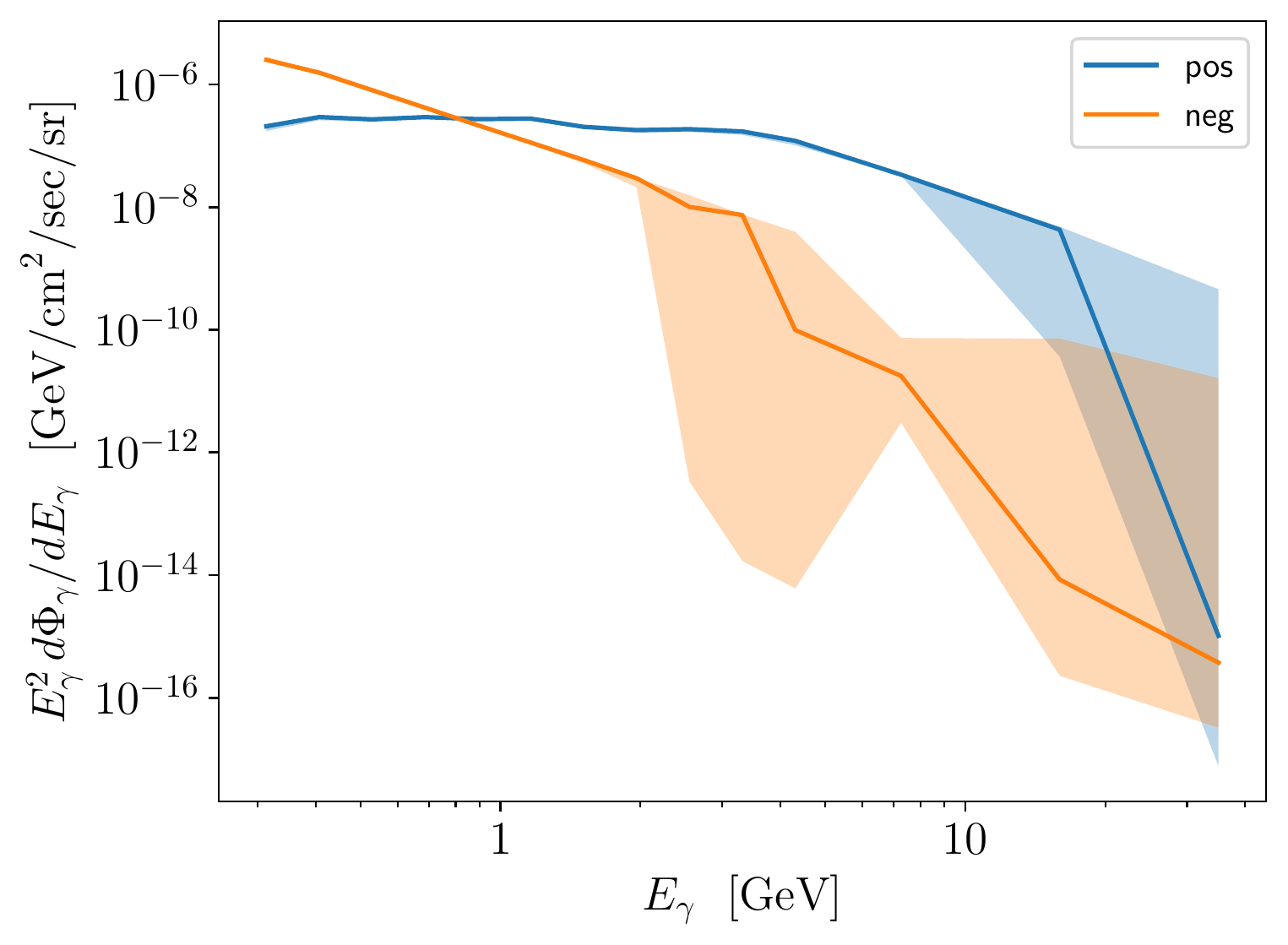}~~
\includegraphics[width=0.45\textwidth]{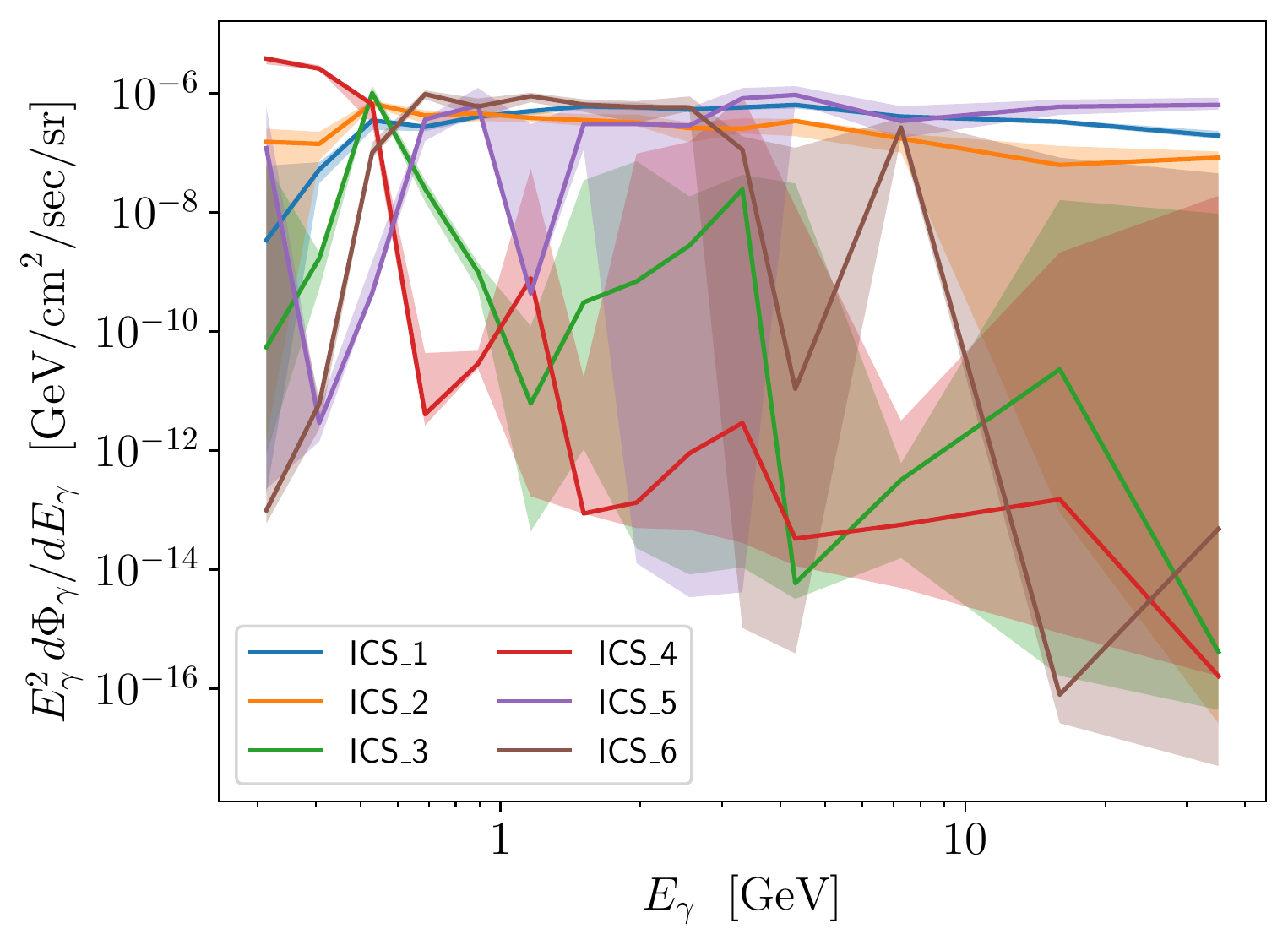}
\caption{Best-fit spectra and 95\% credible intervals of the flux of the ring-based templates that were fit alongside the boxy bulge excess template. For the negative residual component, we show its absolute value in the lower left panel.}
\label{fig:spectra_r}
\end{center}
\end{figure*}
In Fig.~\ref{fig:spectra_r}, we show the spectra of the best-fit ring-based templates that were fit alongside the boxy bulge, which was the excess with the relative best fit conditioned on using ring-based templates. In the four panels, clockwise from top left, we show the {\sf HI} fluxes, the {\sf H2}  fluxes, the {\sf ICS} fluxes, and the ``residual'' fluxes. As is apparent, the dominant contributions of the {\sf HI},  {\sf H2},  and ``residual'' fluxes have smooth, power-law like behavior. The subdominant {\sf HI} and {\sf H2}  fluxes, which account for of order tens of percent of the emission of these components, show sharp, non-monotonic behavior between energy bins, which is very difficult to explain with any physical model. The sum of the absolute value of the ``residual'' components comprise tens of percents of the emission at low energies. 

Most strikingly, the {\sf ICS} fluxes are extremely non-smooth in all energy bins: five of the six different rings come to dominate in at least one bin over the fourteen energy bins we fit. Four of the six rings have extremely jagged, non-smooth behavior. The two innermost rings, labelled {\sf ICS\_1} and {\sf ICS\_2} by the convention of \cite{Pohl:2022hydrogen}, which we follow, are somewhat smoother than the other {\sf ICS} rings.

~\\ \noindent {\bf Results with Macias et al.~maps:}
In Tab.~\ref{tab:resultsM} we report the results of performing the fits as described in the main text, but using the sixteen ring-based maps from \cite{Macias:2016nev, Macias:2019omb, Abazajian:2020tww} instead. Compared to the results in the main text using the ring-based maps of \cite{Pohl:2022hydrogen}, the quality of the fits is substantially worse for every model of the excess. In addition, we observe that the ``residual'' components dominate the emission unless strict priors are imposed.

\setlength{\tabcolsep}{1pt}
\begin{table}[t]
\caption{Comparison of the templates of \cite{Macias:2016nev, Macias:2019omb, Abazajian:2020tww} to the templates from \cite{Pohl:2022hydrogen}, as in Tab.~\ref{tab:results}.}
\begin{center}
\begin{tabular}{c|l|l}
Excess Model & $-2 \, \Delta \! \ln \cL$ & $\Delta \!\ln \mathcal B$ \\ \hline
No Excess & $+9828$ & $-4790$ \\
X-Shaped Bulge & $+9851$ & $-4901$ \\
Dark Matter & $+9512$ & $-4643$ \\
Boxy + X-Shaped Bulges & $+7808$ & $-3778$ \\
Boxy Bulge & $+7805$ & $-3897$ \\
Boxy Bulge ``plus'' & $+8026$ & $-3907$
\end{tabular}
\end{center}
\label{tab:resultsM}
\end{table} %

~\\ \noindent {\bf Implementing %entirely agnostic
wider priors on the template normalizations:}
In the main text we showed that for reasonable choices on the normalizations of the 
templates used to describe the galactic diffuse emission components, we consistently get a clear 
preference for Dark Matter annihilation-like morphology for the GCE over those predicted for the Boxy Bulge
\cite{2002A&A...384..112L, Macias:2016nev, Bartels:2017vsx} or the X-Shaped Bulge \cite{Macias:2016nev, 
Bartels:2017vsx, Macias:2019omb}. In evaluating the results shown in Tab.~\ref{tab:results}, we used for the 
HI, H2, and ICS ring-templates, normalization values (independent at every energy bin) that were in the 
range of $[10^{-2}, 10^{10}]$ relative to the energy-independent templates produced by \cite{2002A&A...384..112L, Macias:2016nev, Bartels:2017vsx, Macias:2016nev, Bartels:2017vsx, Macias:2019omb}. For the positive residual template, 
the normalization values were allowed to float in the range of $[10^{-2}, 10^6]$, while for the negative residual template, the 
normalization values were $[10^{-2}, 10^4]$: this restricted range was chosen to ensure that no parameter points in any energy bin or any pixel could result in a negative flux expectation value, in spite of the negative value of the negative residual. Finally, for the bubbles, isotropic, and GCE templates, the normalization values were 
allowed to float in the range of $[10^{-2}, 10]$ relative to their values in \cite{Cholis:2021rpp}.

In this section we allow our priors to have wider ranges.
We note that there is no good justification for such a choice from the point of astrophysical processes in the 
Milky Way%. Moreover, as we show some of the resulting fluxes from that fitting come with amplitudes entirely 
, and the converged results lead to preferred fluxes that we consider
unjustifiable. However, even with the most arbitrary normalizations in the fitting procedure the basic results of our work are still valid.
Those are, i) that using the templates of Ref.~\cite{Cholis:2021rpp} is preferred to the templates of Ref.~\cite{Pohl:2022hydrogen} 
and ii) that the Dark Matter annihilation-like morphology is preferred to the X-Shaped Bulge or the Boxy Bulge.

In Tab.~\ref{tab:resultsWP}, we perform the same comparison as in Tab.~\ref{tab:results} but with the unphysically wide priors described here.
As with the previous tables, %a negative value in the last column indicates statistical preference in this case for the AP over the RB backgrounds.  
a lower $-2\Delta \!\ln \mathcal L$ and a higher $\Delta \!\ln \mathcal B$ indicate a preference for a given model.
We note that as with Tab.~\ref{tab:results}, we have found that the Dark Matter annihilation-like template combined with the Boxy 
Bulge template (with free relative normalizations)
is slightly preferred for the GCE component as compared to a pure Dark Matter annihilation-like morphology,
and both of these are substantially favored over any alternative. 

%\begin{table}
%\caption{Results with agnostic priors on the templates normalizations. \textcolor{blue}{\bf{sdm: I think we should write in the same way as table 1 -- otherwise, it's hard to interpret and maybe (to a conspiracy minded person) kinda seems like we're hiding something. Also, it makes it a bit difficult to compare those results and these}}}
%\begin{center}
%\begin{tabular}{c|c|c}
%Excess Model & Bgd.~Templates & $-2 \Delta \! \ln \cL$ \\ 
%No Excess & AP - RB & 1805 \\
%X-Shaped Bulge & AP - RB & $574$  \\
%Boxy Bulge & AP - RB & $-52$  \\ 
%Boxy Bulge ``plus'' & AP - RB & $-131$  \\ 
%Dark Matter & AP - RB & $-942$  \\
%DM + Boxy Bulge ``plus'' & AP - RB & $-1056$  \\ \hline
%\end{tabular}
%\end{center}
%\label{tab:resultsWP}
%\end{table}%
\begin{table}[t]
\caption{Alternative comparison of models of the GCE, similar to similar to Tab.~\ref{tab:results} but with unphysically wide priors. The first seven results, generated in this work, rely on the ring-based method of \cite{Pohl:2022hydrogen} to describe astrophysical emission. The final five results utilize templates from \cite{Cholis:2021rpp}.}
\begin{center}
\begin{tabular}{c|c|l|l}
Excess Model & Bgd.~Templates & $-2 \Delta \! \ln \cL$ & $\Delta \!\ln \mathcal B$ \\ \hline
No Excess & ring-based \cite{Pohl:2022hydrogen} & 0 & 0 \\
X-Shaped Bulge & ring-based \cite{Pohl:2022hydrogen} & $-16$ & $-115$ \\
Dark Matter & ring-based \cite{Pohl:2022hydrogen} & $-542$ & $+251$ \\
Boxy \& X-Shaped Bulges & ring-based \cite{Pohl:2022hydrogen} & $-350$ & $+119$ \\
Boxy Bulge & ring-based \cite{Pohl:2022hydrogen} & $-414$ & $+142$ \\
Boxy Bulge ``plus'' & ring-based \cite{Pohl:2022hydrogen} & $-466$ & $+156$ \\ 
Boxy Bulge ``plus'' \& DM & ring-based \cite{Pohl:2022hydrogen}  & $-734$ & $+351$ \\ \hline
No Excess & astrophysical \cite{Cholis:2021rpp} & $+1805$ & $-50$ \\
Boxy Bulge & astrophysical \cite{Cholis:2021rpp} & $-53$ & $+835$ \\ 
Boxy Bulge ``plus'' & astrophysical \cite{Cholis:2021rpp} & $-132$ & $+875$ \\ 
Dark Matter & astrophysical \cite{Cholis:2021rpp} & $-943$ & $+1290$ \\
Boxy Bulge ``plus'' \& DM & astrophysical \cite{Cholis:2021rpp} & $-1056$ & $+1320$
\end{tabular}
\end{center}
\label{tab:resultsWP}
\end{table}%

\begin{figure}[t]
\begin{center}
\includegraphics[width=0.45\textwidth]{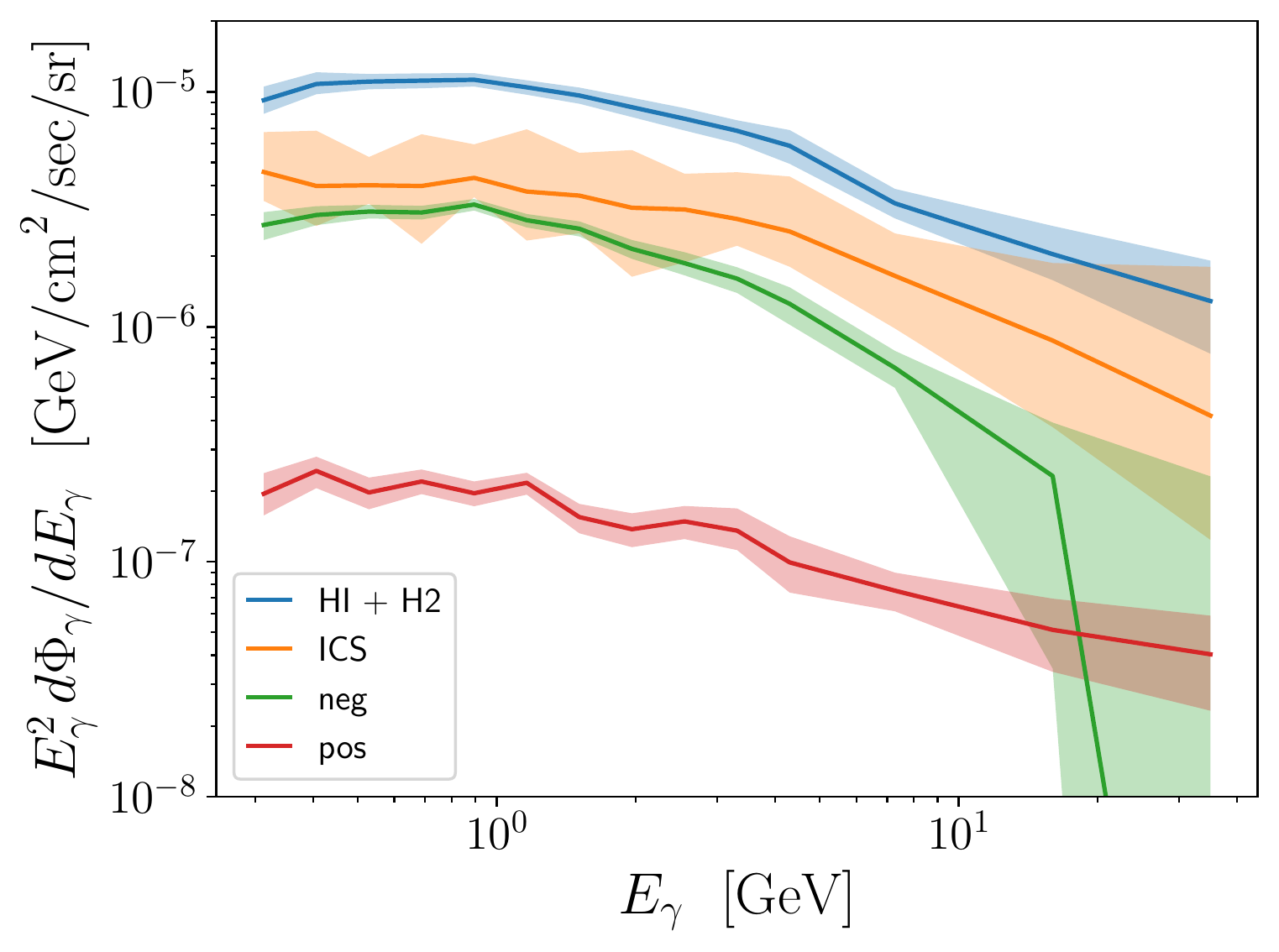}
\caption{Best-fit spectra and 95\% credible intervals for the background components using the ring-based templates of \cite{Pohl:2022hydrogen}.}
\label{fig:spectra_WP}
\end{center}
\end{figure}

In Fig.~\ref{fig:spectra_WP}, using the ring-based templates of Ref.~\cite{Pohl:2022hydrogen}, we show the best-fit spectra and 
95\% credible intervals for the background components. We sum all the gas HI and H2 template fluxes (eight templates) into the blue 
line and band. The combined six ICS templates are shown in orange. The ``positive residual'' gas template is given in red and the (absolute value of the) ``negative residual'' gas template   
in green. The need for a positive residual template can be attributed to unmodeled absorption to the relevant 21 cm line for HI and 
the CO 2.6 mm line used to model the H2 distribution\footnote{The models of \cite{Cholis:2021rpp} include a wide variety of assumptions on the on the distribution of HI and H2 gasses. They do that by testing alternative choices on the absorption of the HI 21cm line and the 
conversion from the CO line observations to the H2 distribution. They also use complementary observations of the cosmic dust emission, that can be 
used as an alternative probe for the ISM gas distribution. Thus, the models of \cite{Cholis:2021rpp} include these contributions self-consistently.}. However, the negative residual template can not be attributed to any 
unmodeled absorption of the HI or CO lines. The motivation behind a negative residual gas template is not clear. We note that the positive 
residual gas map provides a $\sim 2\%$ correction to the combined HI + H2 templates, something that could be expected for the $40^{\circ} \times 
40^{\circ}$ ROI. However, the negative residual gas map provides a $\sim 30\%$ correction to the combined HI + H2 templates, when even 
its existence is not well founded. We consider such a result an alarming indication of the possible self-consistency of ring-based templates with arbitrary normalizations, and restrict to a narrower range of the ``residual'' maps for the purpose of reporting our main results.

\end{appendix}  
                  
\bibliography{GCE_bib}

\end{document}